\documentclass[twocolumn]{aastex631}

\usepackage{hyperref}

\shorttitle{SN~2023ixf in M101}
\shortauthors{Teja et al.}

\begin{document}

\title{Far-Ultraviolet to Near-Infrared Observations of SN~2023ixf: A high energy explosion engulfed in complex circumstellar material}

\correspondingauthor{Rishabh Singh Teja, Avinash Singh}
\email{rishabh.teja@iiap.res.in, avinash21292@gmail.com }

\author[0000-0002-0525-0872]{Rishabh Singh Teja}
\affiliation{Indian Institute of Astrophysics, II Block, Koramangala, Bengaluru-560034, Karnataka, India}
\affiliation{Pondicherry University, R.V. Nagar, Kalapet, Pondicherry-605014, UT of Puducherry, India}

\author[0000-0003-2091-622X]{Avinash Singh}
\affiliation{Hiroshima Astrophysical Science Center, Hiroshima University, Higashi-Hiroshima, Hiroshima 739-8526, Japan}

\author[0000-0001-7570-545X]{Judhajeet Basu}
\affiliation{Indian Institute of Astrophysics, II Block, Koramangala, Bengaluru-560034, Karnataka, India}
\affiliation{Pondicherry University, R.V. Nagar, Kalapet, Pondicherry-605014, UT of Puducherry, India}

\author[0000-0003-3533-7183]{G.C. Anupama}
\affiliation{Indian Institute of Astrophysics, II Block, Koramangala, Bengaluru-560034, Karnataka, India}

\author[0000-0002-6688-0800]{D.K. Sahu}
\affiliation{Indian Institute of Astrophysics, II Block, Koramangala, Bengaluru-560034, Karnataka, India}

\author[0000-0002-7708-3831]{Anirban Dutta}
\affiliation{Indian Institute of Astrophysics, II Block, Koramangala, Bengaluru-560034, Karnataka, India}
\affiliation{Pondicherry University, R.V. Nagar, Kalapet, Pondicherry-605014, UT of Puducherry, India}

\author[0000-0002-7942-8477]{Vishwajeet Swain}
\affiliation{Department of Physics, Indian Institute of Technology Bombay, Powai, Mumbai 400076}

\author{Tatsuya Nakaoka}
\affiliation{Hiroshima Astrophysical Science Center, Hiroshima University, Higashi-Hiroshima, Hiroshima 739-8526, Japan}

\author[0009-0002-7897-6110]{Utkarsh Pathak}
\affiliation{Department of Physics, Indian Institute of Technology Bombay, Powai, Mumbai 400076}

\author[0000-0002-6112-7609]{Varun Bhalerao}
\affiliation{Department of Physics, Indian Institute of Technology Bombay, Powai, Mumbai 400076}

\author[0000-0002-3927-5402]{Sudhanshu Barway}
\affiliation{Indian Institute of Astrophysics, II Block, Koramangala, Bengaluru-560034, Karnataka, India}

\author[0000-0003-0871-4641]{Harsh Kumar}
\affiliation{Department of Physics, Indian Institute of Technology Bombay, Powai, Mumbai 400076}

\author[0000-0002-8070-5400]{Nayana A.J.}
\affiliation{Indian Institute of Astrophysics, II Block, Koramangala, Bengaluru-560034, Karnataka, India}

\author{Ryo Imazawa}
\affiliation{Department of Physics, Graduate School of Advanced Science and Engineering, Hiroshima University, Kagamiyama, 1-3-1 Higashi-Hiroshima, Hiroshima 739-8526, Japan}

\author[0000-0001-7225-2475]{Brajesh Kumar}
\affiliation{Aryabhatta Research Institute of Observational Sciences, Manora Peak, Nainital-263001, Uttarakhand, India}

\author[0000-0001-6099-9539]{Koji S. Kawabata}
\affiliation{Hiroshima Astrophysical Science Center, Hiroshima University, Higashi-Hiroshima, Hiroshima 739-8526, Japan}

\begin{abstract}

We present early-phase panchromatic photometric and spectroscopic coverage spanning far-ultraviolet (FUV) to the near-infrared (NIR) regime of the nearest hydrogen-rich core-collapse supernova in the last 25 years, SN~2023ixf. We observe early `flash' features in the optical spectra due to a confined dense circumstellar material (CSM). We observe high-ionization absorption lines (\ion{Fe}{2}, \ion{Mg}{2}) in the ultraviolet spectra from very early on. We also observe a multi-peaked emission profile of H$\rm\,\alpha$ in the spectrum beginning $\sim 16$~d, which indicates ongoing interaction of the SN ejecta with a pre-existing shell-shaped CSM having an inner radius of $\sim$\,75 AU and an outer radius of $\sim$\,140 AU. The shell-shaped CSM is likely a result of enhanced mass loss $\sim$\,35 - 65 years before the explosion assuming a standard Red-Supergiant wind. The UV spectra are dominated by multiple highly ionized narrow absorption features and broad emission features from elements such as C, N, O, Si, Fe, and Ni. Based on early light curve models of Type II SNe, we infer that the nearby dense CSM confined to $\rm 7\pm3\times10^{14}~cm$ ($\sim 45$~AU) is a result of enhanced mass loss ($\rm 10^{-3.0\pm0.5}~M_\odot~yr^{-1}$) two decades before the explosion.

\end{abstract}


\section{Introduction}
\label{sec:intro}

Massive stars ($\rm \gtrsim 8~M_\odot$) that meet their fate with the explosive phenomena are termed as Core-collapse Supernovae (CCSNe). CCSNe are either hydrogen-rich (Type II) or hydrogen-poor (Ib, Ic) \citep{1997Filippenko}. Recent advancements in all-sky surveys (e.g., ZTF, ATLAS) have made it possible to discover young supernovae when rapid changes occur in their light curves, spectral energy distribution, and spectral evolution apart from increasing brightness \citep{2016ApJ...818....3K18p, 2022arXiv221203313B36}. The early evolution of a good fraction ($>36\%$) of Type II SNe is dominated by narrow emission features associated with confined dense circumstellar material (CSM, \citealp{2021ApJ...912...46B30,2022arXiv221203313B36}). The characteristics of the nearby dense CSM are visible in the spectral sequence as ``flash" features consisting of narrow high-ionization lines that last a few to several days depending on the radius and density of the CSM \citep{2014Natur.509..471G, 2017yaron, 2022ApJ...924...15J}. The flash features are caused by the ionizing photons that result when the shock breaks out from the stellar surface and flashes/ionizes the nearby circumstellar material. Some authors \citep{2019MNRAS.483.3762K, 2022ApJ...924...15J} have noted that the ionization from shock breakout lasts for a few hours only, and to get prolonged flash features, another photon source is required, such as ejecta-CSM interaction. The earliest detailed time series observations of ``flash spectroscopy'' were observed for SN~2013fs \citep{2017yaron}. The confined CSM ($<10^{15}$ ~cm) of SN~2013fs was indicated by the disappearance of flash features, and it was consistent with the radio non-detection \citep{2017yaron}. It was argued that this could only result if the progenitor had undergone a short-lived episode of enhanced mass loss just a few years before the explosion. Even though the rise of all-sky surveys has led to an order-of-magnitude increase in the early detection (and follow-up) of such events \citep{2018PASP..130c5003B, 2021A&G....62.5.34N}, the physics behind the specifics of such interaction and origins of CSM are still not definitively understood, and the associated observables such as light curves are not very well constrained \citep{2017MNRAS.470.1642F, 2021ApJ...906....3W, 2022A&A...660L...9D, 2022ApJ...930..168K, 2023PASJ...75..634M}.

The CCSNe have been studied extensively in optical and near-infrared (NIR) regimes, but extensive studies in the ultraviolet (UV) regime are still limited \citep{2014ApJ...787..157P, 2007ApJ...659.1488B, 2023arXiv230406147V}. The crucial aspect of the observational investigation in UV is the requirement of observation from space-based missions \citep{2022ApJ...934..134V, 2023arXiv230501654B} for which scheduling time disruptive Target-of-Opportunity (ToO) observations is not rapid for a majority of missions. The flux in UV declines very rapidly, requiring prompt observations and follow-ups. The UV emission from young CCSNe allows the investigation of hot and dense ejecta and/or the presence of CSM when the photosphere is located in the outer layers of the progenitor star \citep{2009bufano}. Many Type II SNe show a nearly featureless early optical spectral sequence, unlike the far-UV and near-UV, which showcases a plethora of metal features. The detection of these features can be used to determine the composition of the outer envelope of the pre-SN star, the temperature of the outer layers of the ejecta, or the CSM and its characteristics \citep{2022dessart, 2023arXiv230501654B}.

SN~2023ixf was discovered on 2023 May 19 17:27:15.00 UT (JD~2460084.23) in the galaxy M~101 at $\sim 14.9$~mag in `clear' filter \citep{2023TNSTR1158....1I} and classified as a Type II SN \citep{2023perley, 2023TNSCR1233....1T}. The pre-discovery photometry from Zwicky Transient Facility (ZTF) and other Transient Name Server (TNS) alerts provide tight constraints on the explosion epoch. Using the last non-detection (JD~2460083.31) and first detection (JD~2460083.32) \citep{2023TNSAN150}, we find the explosion epoch, $\rm t_{exp} = JD~2460083.315\pm0.005$ which has been used throughout this work. We note that the last-non detection used is not very deep ($>$\,18 mag) and if we consider the deeper non-detection \citep[$>20.5$\,mag,][]{2023TNSAN130} on JD~2460083.16, the explosion epoch has a marginal change (of $\sim$\,0.08~d) to JD~24600083.235.

Several professional and amateur astronomers have followed up on SN~2023ixf, being one of the nearest CCSNe in the last 25 years. Various time-domain groups across the globe have been monitoring it soon after its discovery, and the results based on the early observations have already been presented. The early phase optical and NIR photometry and optical spectroscopy have been presented by \citet{2023arXiv230600263Y, 2023arXiv230606097H, 2023arXiv230604721J}. The  presence of flash features in the spectra and increased luminosity is interpreted as due to the presence of nitrogen/helium-rich dense CSM and the interaction of supernova ejecta with it \citep{2023arXiv230600263Y, 2023arXiv230604721J}. By comparing the early light curve with the shock cooling emission \citep{2023arXiv230606097H} suggested that the progenitor of SN 2023ixf could be a red supergiant with radius 410$\pm$ 10 R$_\odot$. The high-resolution spectroscopy revealed that the confined CSM is asymmetric \citep{2023arXiv230607964S}. Pre-imaging data was analyzed at the SN~2023ixf site in recent works, constraining the  mass of the progenitor between $12-17$~M$_\odot$ \citep{2023ApJ...952L..30J, 2023arXiv230514447P, 2023TNSAN.139....1S}. These estimates are well within the earlier detected CCSNe progenitors \citep{2009ARA&A..47...63S, 2017RSPTA.37560277V}.
 
This letter presents the panchromatic evolution of SN~2023ixf spanning the far-ultraviolet (FUV) to near-infrared (NIR) wavelengths during the first three weeks since its discovery. The flow of the paper is as follows: In Section~\ref{sec:obsdata}, we estimate the distance to the host galaxy and its extinction and briefly describe the source of data acquisition and the reduction procedure. Further, we present our spectroscopic observation in Section~\ref{sec:spectra} along with its analysis and modeling in different regimes. Later on in Section~\ref{sec:lightcurves}, we describe the light curve evolution and its early phase analysis. We summarize and discuss this early phase work in Section~\ref{sec:discussion}. 

\section{Observations and Data Reduction}
\label{sec:obsdata}

SN~2023ixf exploded in the outer spiral arm of the host galaxy, M~101, a face-on giant spiral galaxy that lies comparatively close to the Local Group. \citet{2015trgb} estimated a mean distance of 6.79\,$\pm$\,0.14 Mpc ($\rm \mu$ = 29.15\,$\pm$\,0.05 mag) to M~101 using the tip of the RGB method \citep{1993lee} with low-uncertainty. \citet{2022riess} used Cepheids to estimate a distance of 6.85\,$\pm$\,0.15 Mpc ($\rm \mu$ = 29.18\,$\pm$\,0.04 mag). We adopt a mean distance of 6.82\,$\pm$\,0.14 Mpc ($\rm \mu$ = 29.17\,$\pm$\,0.04 mag). The gas phase metallicity was computed by \citet{2022garnerm101} using various \ion{H}{2} regions in the galaxy and estimated an oxygen abundance of 12 + log[O/H]\,$\sim\,$ 8.7 in the outer spiral arms of the galaxy which is similar to solar metallicity \citep{2009asplund}. 

Galactic reddening in the line-of-sight of SN~2023ixf inferred from the dust-extinction map of \citet{2011schlafly} is $E(B-V)$\,=\,0.0077$\pm$0.0002 mag. Using high-resolution data, \citet{2023TNSAN.160....1L} computed equivalent widths of $\rm Na\,$I~D1 and D2 lines to be 0.118 \AA\ and 0.169 \AA, respectively. Using the relation from \citet{2012poznanski}, we infer a mean host reddening of $E(B-V)$\,=\,0.031 $\pm$ 0.011 mag using $\rm Na\,$I~D1 and D2. A total reddening of $E(B-V)$\,=\,0.039\,$\pm$\,0.011 mag is adopted for SN~2023ixf which is consistent with \citet{2023arXiv230607964S}.
 
\subsection{Optical and Near-Infrared}

We carried out broadband optical photometric observations in SDSS $u'g'r'i'z'$ filters beginning 2023 May 20 UT, using the robotic 0.7-m GROWTH-India telescope (GIT, \citealp{2022growth}) located at Indian Astronomical Observatory (IAO), Hanle, India. Data were downloaded and processed with the standard GIT image processing pipeline described in \citet{2022growth}. While standard processing was sufficient for $g'r'i'z'$ bands, the $u'$ band data did not have enough stars for automated astrometry using \texttt{astrometry.net} \citep{2010AJ....139.1782L} and further zero-point estimation. The zero point was computed manually using several non-variable SDSS stars available in the SN field for GIT images. Optical spectroscopic observations of SN~2023ixf were carried out using the HFOSC instrument mounted on the 2-m Himalayan Chandra Telescope (HCT), IAO \citep{2014Prabhu}. The spectroscopic data were reduced in a standard manner using the packages and tasks in \texttt{IRAF} \citep[For details refer,][]{2023arXiv230610136T}.

Near-Infrared (NIR) data were obtained from the Hiroshima Optical and Near-InfraRed Camera \citep[HONIR;][]{2014akitaya} mounted at the 1.5-m Kanata Telescope. The NIR data were reduced using standard procedures in \texttt{IRAF}, and the calibration was done using secondary stars from the 2MASS catalog \citep{2006AJ....131.1163S}.

\subsection{Ultraviolet}

SN~2023ixf was observed by the UltraViolet Imaging Telescope (UVIT; \citealt{Kumar+2012, Tandon+2017}) on board \textit{AstroSat} on 2023 May 25 \& 30 UT in both imaging and spectroscopic modes. However, we could only use imaging data from May 30 for photometry since the images from the earlier epoch are saturated. The spectra obtained at all epochs are of good quality and have been used for this study. We also triggered the \textit{UVIT} for several Target of Opportunity (ToO) proposals. But due to technical constraints, observations against our ToO request could be undertaken only on June 11, 2023. However, data obtained through ToO observations are immediately made public at the Indian Space Science Data Center (ISSDC) portal\footnote{\href{https://www.issdc.gov.in/astro.html}{ISSDC Portal}}, and we have used the Level 1 (raw) and Level 2 (processed) data files available at ISSDC in this work. All the UVIT observations are listed in Table~\ref{UVIT_obs_log}. \textit{UVIT} observations were performed with the FUV \textit{F172M} and \textit{F148W} filters and with the FUV gratings \textit{Grating1} and \textit{Grating2}. These two gratings are mounted on the FUV filter wheel at positions F4 and F6, respectively \citep{Kumar+2012} and have perpendicular dispersion axes. The \textit{AstroSat-UVIT} data were pre-processed with \texttt{CCDLAB} \citep{2017PASP_postma} following the steps described in \cite{2021_postma}. Aperture photometry was performed using a 12-pixel (5$\arcsec$) aperture and calibrated following the procedures mentioned in \cite{2020_Tandon}. Spectral extraction and calibrations were performed manually following the procedure described in \cite{2020_Tandon} and \cite{2021_Dewangan} using \texttt{IRAF} and \texttt{python}.

SN~2023ixf was also monitored extensively by the Ultraviolet Optical Telescope (UVOT; \citealp{2005roming}) onboard the Neil Gehrels Swift Observatory \citep{2004gehrels} beginning May 21, 2023. We utilize the publicly available data obtained from Swift Archives\footnote{\href{https://www.swift.ac.uk/swift_portal/}{Swift Archive Download Portal}}.  Photometry was performed using the UVOT data analysis software in \texttt{HEASoft}, following the procedure described in \citet{2022rishabh}. To check for contamination, we looked at the archival \textit{Swift} data of the host galaxy M~101 (ID 00032081). The count rates at the SN site for an aperture similar to that used for the SN photometry are insignificant and comparable to the background. Being a very bright SN, most photometric data points were saturated. We checked the \texttt{saturate} and \texttt{sss\_factor} flags from the output and discarded all the saturated and unusable data points based on those flags. Spectroscopic data reduction for \textit{Swift} UV-grism data was performed using the standard \texttt{UVOTPY} package, which includes the latest grism calibrations and corrections \citep{2014ascl.soft10004K}. Further, multiple spectra captured intra-night were summed using \texttt{uvotspec.sum\_PHAfiles} program in \texttt{UVOTPY} to increase the overall SNR. The first two spectra separated by just 0.1~d showed intranight flux variability due to the rapid rise; hence, these two spectra were not summed. Around 1800~\AA, a few spectra were contaminated by a strong source, therefore, we have considered the UVOT spectra beyond 1900~\AA\ only.

\subsection{X-rays}
\label{sec:xrays}

SN~2023ixf was also observed with the Soft X-ray Telescope (SXT) covering the 0.3-7.0 keV energy band \citep{singh_2016,singh_2017} aboard \emph{AstroSat} \citep{singh_2014}. Data were obtained in photon counting (PC) and fast window (FW) modes over multiple orbits starting May 25 (see Table~\ref{UVIT_obs_log}). Orbit-wise Level 2 data were downloaded from ISSDC and merged into a single cleaned event file using the standard \textit{Julia} based merger tool. Images, spectra, and lightcurves were produced using \texttt{XSELECT} v2.5a from \texttt{HEASoft} 6.30.1. We do not obtain a statistically significant detection of the source in the data obtained from SXT observations, possibly due to low exposure times and pointing offsets. However, it was detected by other X-ray facilities, primarily in hard X-rays, with the following reports on  \texttt{ATel}: \emph{NuSTAR} \citep[May 22, ][]{NuSTAR2023-ATel16049}, \emph{ART-XC} \citep[May 26 and 29, ][]{ARTXC2023-ATel16065}, and \emph{Chandra} \citep[May 31, ][]{2023ATel16073....1C}.

\begin{table}[]
    \centering
    \caption{Log of AstroSat observations. }
    \label{UVIT_obs_log}
    \begin{tabular}{lllll}
    \hline
   ObsID & Date & Phase & Instrument & Time \\
    &  & (d) & & (ks) \\
    \hline
    T05\_108T01\_ & 2023-05-25 & +6.9&  UVIT FUV & 7.32 \\
    9000005664 &  &  & SXT (FW) & 7.90 \\
    T05\_110T01\_ & 2023-05-30 & +11.9&  UVIT FUV & 4.32 \\
    9000005672 &  &  & SXT (PC) & 8.24 \\
    T05\_116T01\_ & 2023-06-11 & +23.4 &  UVIT FUV  &  3.48 \\
    9000005682$^a$ & &  &  SXT (PC) & 15.24 \\
    \hline
    \end{tabular}
    \\

 $^a$ Observation against our ToO
    
\end{table}

\subsection{Other Data Sources}

Being a nearby SN in one of the most well-observed host galaxies, M~101, many amateur astronomers and professional observatories have monitored the SN. We supplemented our photometric dataset with various detections and non-detections of SN~2023ixf from Astronomer's Telegrams\footnote{\href{https://www.astronomerstelegram.org}{Astronomer's Telegrams}} and TNS Astronotes\footnote{\href{https://www.wis-tns.org/astronotes}{TNS AstroNote}}, and include the magnitudes reported by \citet{2023TNSAN120, 2023TNSAN123, 2023TNSAN124, 2023TNSAN125, 2023TNSAN128, 2023TNSAN129,2023TNSAN130, 2023TNSAN136, 2023TNSAN141, 2023TNSAN142, 2023TNSAN143, 2023TNSAN144, 2023TNSAN150, 2023TNSAN153, 2023TNSAN154, 2023TNSAN156, 2023TNSAN161, 2023ATel16054....1S}.

\section{Spectral Analysis}
\label{sec:spectra}

\subsection{Optical Spectra}

\begin{figure*}[hbt!]
	 \resizebox{\hsize}{!}{\includegraphics{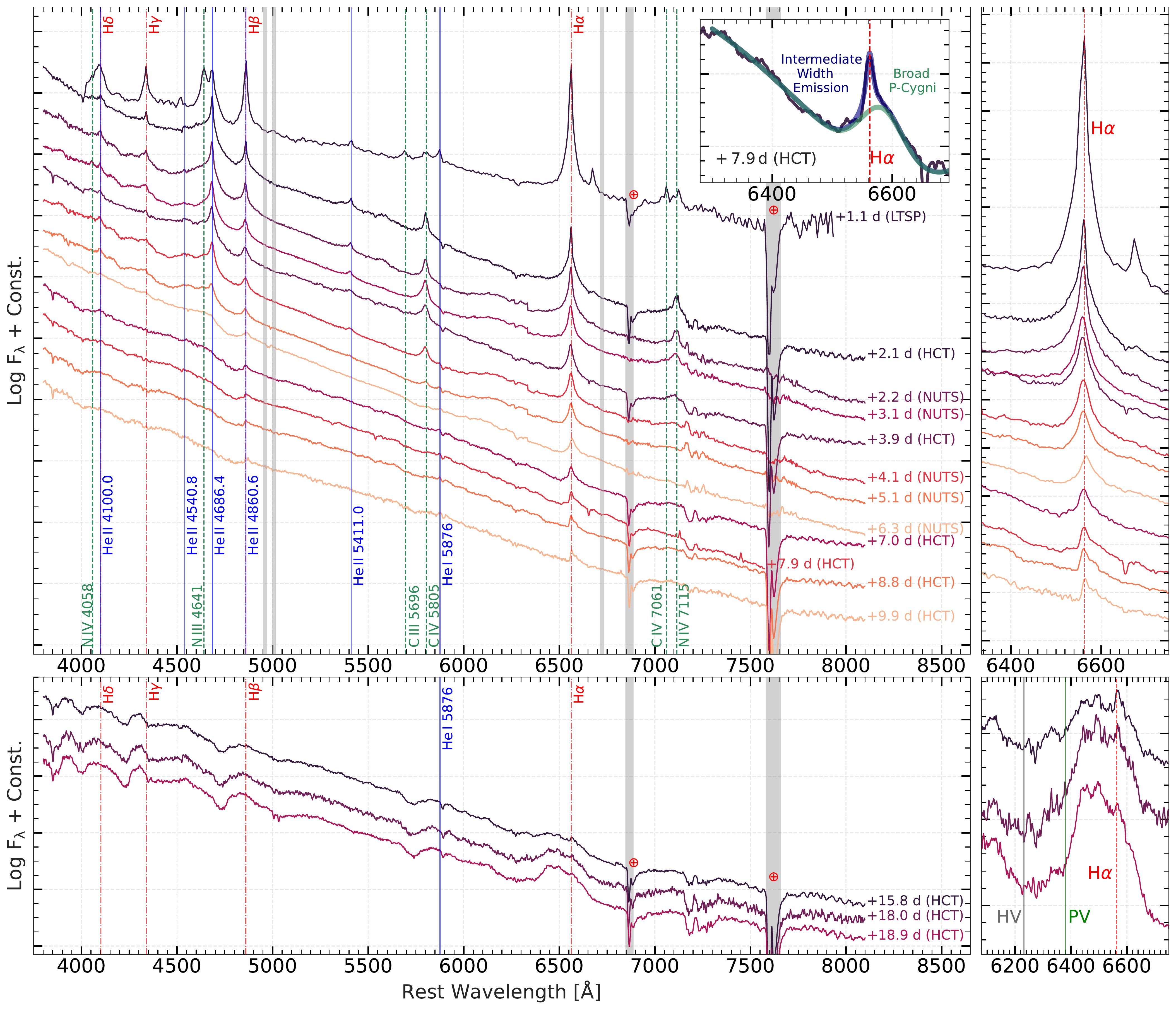}}
    \caption{Optical spectral evolution for SN~2023ixf from HCT, \citet{2023perley} and \citet{2023stritzinger}. The spectra are corrected for the redshift of the host galaxy M~101, and the epochs are labeled with respect to our adopted explosion epoch. {\bf Top: }{\it Left:} Early time spectral sequence of flash features in SN~2023ixf with line identification of high-ionization features and Balmer lines. The inset depicts the H$\rm\,\alpha$ profile on $+$7.9\,d having a broad P-Cygni feature and an Intermediate-width Lorentzian emission. {\it Right:} Evolution of line-profile of H$\rm\,alpha$ during the flash phase. {\bf Bottom: }{\it Left:} Spectral sequence of SN~2023ixf during the photospheric phase. {\it Right:} Evolution of multi-peaked emission profile of H$\rm\,alpha$ during the photospheric phase. HV and PV refer to the high-velocity and photospheric velocity components in the blue-shifted absorption wing of H$\rm\,\alpha$.}
    \label{fig:flashSpectra}
\end{figure*}

The first optical spectrum of SN~2023ixf was obtained within 5 hrs of discovery by the Liverpool Telescope \citep{2023perley}. Our spectroscopic follow-up with HCT began $\sim 2$ days after the explosion. We present the spectral data obtained from HCT  until $\sim 19$ days after the explosion. The spectral sequence is shown in Figure~\ref{fig:flashSpectra}. The early spectra, until $\sim$\,10 d, show a prominent blue continuum with strong high-ionization emission features due to \ion{C}{4}, \ion{N}{4} and \ion{He}{2}, specifically, \ion{C}{4}~5805~\AA, \ion{C}{4}~7061~\AA, \ion{N}{4}~7115~\AA, \ion{He}{2}~4540~\AA, \ion{He}{2}~4686~\AA\ and \ion{He}{2}~5411~\AA\ along with the Balmer lines $\rm H\alpha$, $\rm H\beta$, $\rm H\gamma$, and $\rm H\delta$. Weak signatures of \ion{C}{3}~5696~\AA\ and \ion{N}{3}~4641~\AA\ and \ion{He}{1}~5876~\AA\ are also seen in the spectra. The highly ionized emission features at $\sim$\,2.1\,d are well reproduced by a combination of a narrow Lorentzian (limited by the resolution) and an intermediate-width Lorentzian of 2500 $\rm km\ s^{-1}$. Our findings during the flash-ionization phase are similar to those reported in \citet{2023arXiv230600263Y, 2023arXiv230604721J,2023arXiv230607964S,2023bostroem}.

The strength of the narrow component fades gradually, in contrast to the intermediate width component, as the SN flux rises in the optical wavelengths. Most of the flash features in our spectral sequence disappear after $+7$\,d. In the spectrum of $7.9$\,d, we observe an intermediate-width $\rm H\alpha$ emission at $\sim$\, 1,000 $\rm km\ s^{-1}$ in addition to the emergence of a broad P-Cygni feature with absorption trough. This could possibly be due to residual of ongoing interaction with the dense CSM responsible for the flash-ionized phase. A similar profile is also seen for the H$\beta$ line. Beginning $\sim$\,16 d (bottom-right panel in Figure~\ref{fig:flashSpectra}), we observe a blue-shifted multi-peaked emission profile of H\,$\rm \alpha$ with a broad absorption feature, which mimics the profile of a detached atmosphere \citep{1990jeffery}, and is an indication of the fast-moving SN shock encountering a low-density shell-shaped CSM  \citep{2002pooley}. The multi-peaked emission profile seen here is similar to the boxy-emission profile seen during the photospheric phase in SN~2007od \citep{2010andrews}, SN~2016gfy \citep{2019ApJ...882...68Singh} and SN~2016esw \citep{2018bdejaeger}.

We observe two absorption troughs blue-ward of H$\alpha$ at 8,000 $\rm km\ s^{-1}$ (PV; Photospheric velocity) and 15,000 $\rm km\ s^{-1}$ (HV; High-Velocity) in the spectrum of $\sim$\,16 d. The HV feature, labeled \lq\lq Cachito \rq \rq in the literature, could instead be due to the presence of \ion{Si}{2} 6355 \AA\ \citep{Gutierrez2017_TypeIISample} in the blue-wing of H\,$\rm \alpha$. The estimated velocity ($\rm\sim5000~km~s^{-1}$) is lower than the photospheric velocity if the feature is due to \ion{Si}{2.} We also detect an analogous profile bluewards of H\,$\rm\beta$ with a similar velocity as seen in the H$\alpha$ profile, indicating that the feature is likely due to hydrogen only. However, the possibility of \ion{Si}{2} blended with the HV feature of hydrogen can not be ruled out altogether.

We estimated the photospheric velocity using the minima of the absorption trough of H\,$\rm \beta$, H\,$\rm \gamma$ and \ion{He}{1}\,5876\,\AA. Although velocities estimated from \ion{Fe}{2} act as a reliable tracer of photospheric velocities \citep{2005bdessart}, we used H and He line velocities as they fairly resemble the photospheric velocities early in the photospheric phase \citep{2014afaran}. Using the ejecta velocities (PV and HV) estimated above, we compute an inner radius of $\sim$\,75 AU and an outer radius of $\sim$ 140 AU for the shell-shaped CSM encountered by the SN ejecta. Assuming a standard RSG wind velocity of 10 $\rm km\ s^{-1}$ \citep{2014smith}, the progenitor of SN~2023ixf likely experienced this enhanced mass-loss $\sim$\,35 - 65 years before the explosion. If we consider the wind velocity of $\sim 115$~km s$^{-1}$ inferred by \citet{2023arXiv230607964S} using high-resolution optical spectra, we estimate that mass loss episode likely occurred $\sim 3-6$ years before the explosion.

\subsection{UV Spectra}

\label{sec:UVspectra}

\begin{figure*}[hbt!]
\begin{minipage}[c]{.5\hsize}
	 \resizebox{\hsize}{!}{\includegraphics{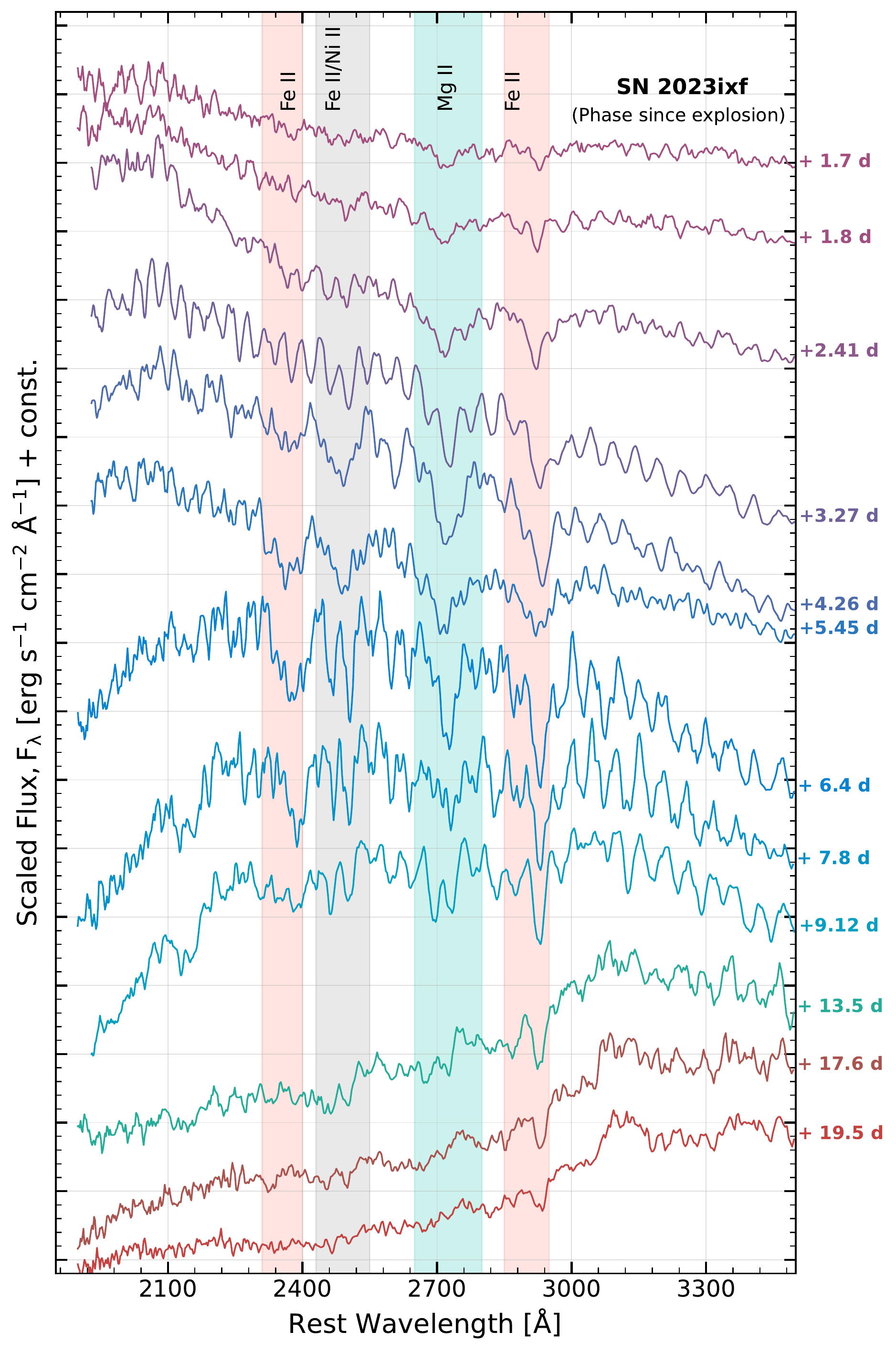}}
     \end{minipage} \hfill
     \begin{minipage}[c]{0.46\hsize}
   \resizebox{\hsize}{!}{\includegraphics{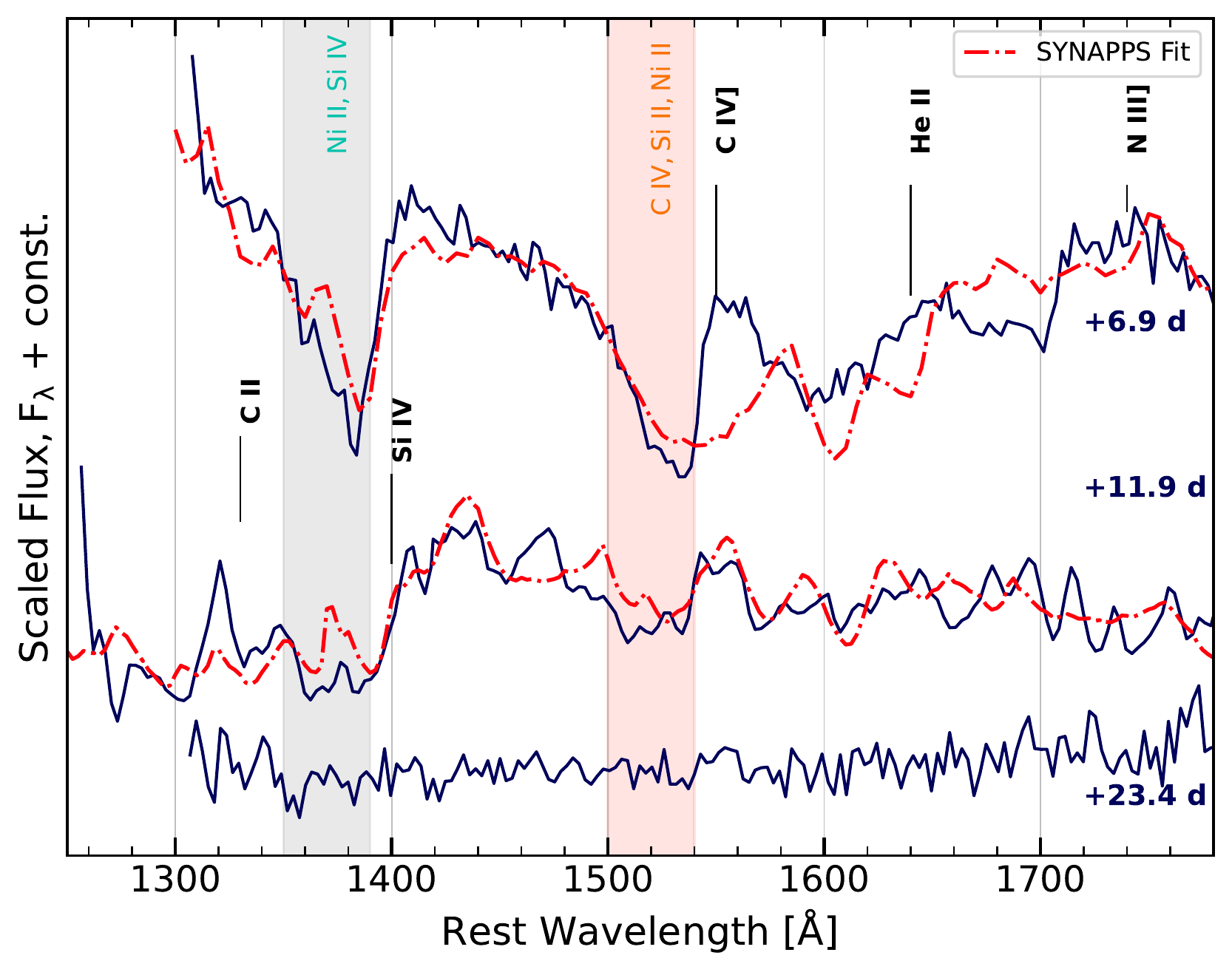}}\\
   \resizebox{\hsize}{!}{\includegraphics{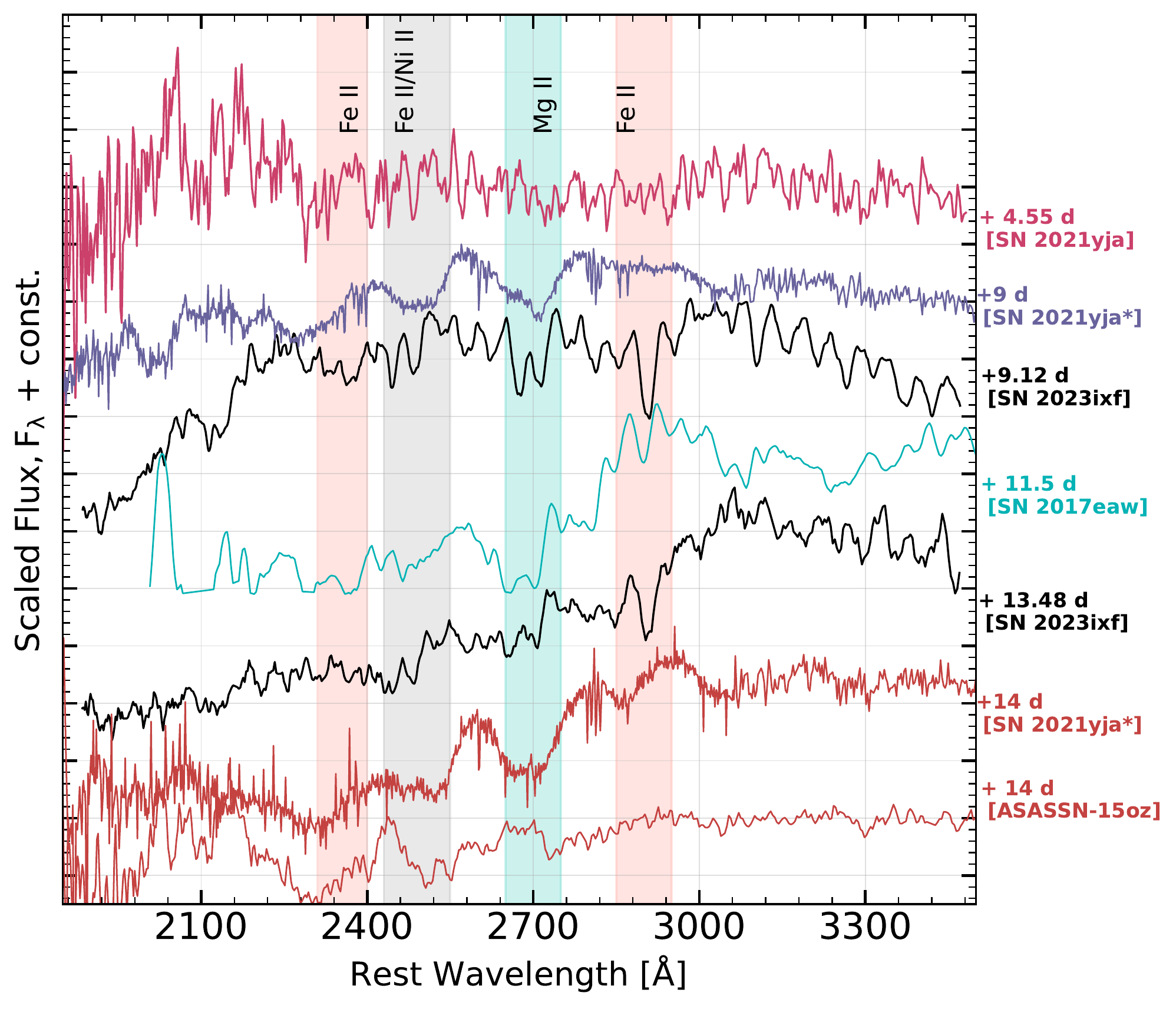}}
  \end{minipage}
    \caption{{\bf Left: } NUV spectral evolution for SN~2023ixf obtained using Swift/UVOT.  {\bf Right: }{\it Top:} FUV spectral evolution obtained using Astrosat/UVIT and the SYNAPPS fit to the spectrum of $\sim$\,7\,d and $\sim$\,12\,d. {\it Bottom:} Spectral comparison of NUV spectra with other Type II SNe. 
   }
    \label{fig:UVSpectra}
\end{figure*}

We present FUV (1250 - 1800 \AA) and NUV (1900 - 3400 \AA) spectral evolution of SN~2023ixf obtained with \textit{AstroSat} and \textit{Swift}, respectively, in Figure~\ref{fig:UVSpectra}. Predominantly, the UV lines arise due to re-emitted UV emission from highly ionized species created from the shock wave expanding into the ambient material \citep{1967ApJ...147..556W, 1981ApJ...251..259C, 1984A&A...133..264F, 1994ApJ...420..268C}. Along with the emission lines, the UV spectra are dominated by a large number of absorption lines from the interstellar matter (ISM) in the Milky Way and the host galaxy due to high ionized states of C, N, O, Si, etc. \citep{1984A&A...133..264F}. Further, the UV spectra are not a simple continuum with isolated emissions and absorptions but a continuous set of features having both emission and absorption features which at times are hard to identify \citep{1995ApJS...99..223P, 2010MNRAS.405.2141D, 2023arXiv230501654B}. The UV spectra of Type II SNe are scarcely studied, particularly the FUV domain, which is largely unexplored. SN~1979C \citep{1980MNRAS.192..861P} was the first Type II SN observed extensively in FUV, and SN~2022acko \citep{2023arXiv230501654B} was the most recent one. For the present work, we restrict ourselves to describing the UV spectra qualitatively.

\subsubsection{FUV spectra}

The FUV spectra of SN~2023ixf were obtained at three epochs $\sim$\,7 d, $\sim$\,12 d, and $\sim\,23$~d (see Table~\ref{UVIT_obs_log}). The first spectrum for SN~2023ixf in FUV is around the optical maximum (Section~\ref{sec:lightcurves}). In the spectrum of $\sim$\,7\,d, we observe two strong absorption bands in the wavelength regions 1340-1400 \AA~ and 1500-1560 \AA~, which can be attributed to a blend of all or potentially a subset of following species \ion{Ni}{2} 1370-1399~\AA, \ion{Si}{4} 1394-1403~\AA~ lines and \ion{C}{4}, \ion{Si}{2} 1527~\AA, \ion{Ni}{2} 1511~\AA~lines, respectively (Figure~\ref{fig:UVSpectra}). Due to the low redshift of SN~2023ixf and with the available spectral resolution, it is difficult to discern whether the interstellar absorptions are either Galactic or due to the host galaxy. We further identify Doppler broadened emission features originating from \ion{C}{4} 1550~\AA, \ion{He}{2}~1640~\AA, and \ion{N}{3}]~1750~\AA~ marked in the top-right panel of Figure~\ref{fig:UVSpectra} similar to SN~1979C \citep{1984A&A...133..264F} and SN~2022acko \citep{2023arXiv230501654B}.

In the spectrum obtained at $\sim$ 12\,d, we continue to observe the two absorption bands but with diminishing depth. Other than the emission features observed in the spectrum of $\sim$\,7\,d, we find emission from \ion{C}{2}~1335~\AA, which could earlier be blended with strong absorption. \ion{Si}{4} and \ion{N}{4}]~ could also be observed in the wavelength region 1400-1500~\AA. As the flux continues to reduce in the FUV region, we see the disappearance of \ion{He}{2} and \ion{N}{3}] emission features. We corroborate the presence of these features by modeling the FUV spectrum at $\sim$\,7\,d and $\sim$\,12\,d using the synthetic spectra generation code \texttt{SYNAPPS} \citep{2011PASP..123..237T}. Many of the features in the spectra could be reproduced in the synthetic spectrum using the high-ionization (up to IV) species of He, C, N, O, S, Si, and Ni. More detailed spectral modeling with multiple elements is required to study these features extensively \citep{2010MNRAS.405.2141D, 2023arXiv230501654B}.  As the SN evolves further, the high density of low-ionization lines of iron-group elements (especially \ion{Fe}{2} and \ion{Fe}{3}) \citep{2000mazzali} amplify the line blanketing in the UV regime as is evident in the FUV spectrum of $\sim 23$~d, which is noisy and featureless owing to the completely extinguished continuum flux. The complete extinction in FUV flux around +20~d is also evident in other Type II objects such as SN~2021yja \citep{2022ApJ...934..134V}, SN~2022wsp \citep{2023arXiv230406147V}, and SN~2022acko \citep{2023arXiv230501654B}.

\subsubsection{NUV spectra}

The first NUV spectrum obtained at $+$\,1.7~d is the earliest-ever NUV spectrum for any CCSN observed after SN~1987A. Contrary to the FUV, many Type II SNe have been observed in NUV at multiple epochs. The NUV spectral coverage of SN~2023ixf is the most comprehensive ever up to +\,20~d after the explosion, with 12 spectra.

We observe weak and blended absorption features in the first spectrum in the wavelength range 2300-3000~\AA. These absorption features continue to grow in strength and width and fully dominate the SN spectra at +\,6.4\,d. The features arise particularly due to \ion{Fe}{2}, \ion{Ni}{2} and \ion{Mg}{2} species \citep{2007ApJ...659.1488B, 2023arXiv230501654B, 2023arXiv230406147V}. The prominence of these absorption features weakens along with increased line blanketing except for the feature present around 2900~\AA, which is observed even in the last spectrum presented here, at +19.5~d.

The flux in NUV started rising from the first epoch and reached a maximum at $\sim$\,5 d after the explosion as the SED transitioned to NUV. In the subsequent epochs, the NUV flux starts declining and drops to the level of the first epoch at around $\sim$\,14 d. There is a significant drop in the flux between +5.5~d and +6.4~d in the region $<$\,2200~\AA, observed with the change in the shape of SED as apparent in the left panel of Figure~\ref{fig:UVSpectra}. This is probably due to the rapid cooling of the SN ejecta coupled with increased line blanketing in the UV wavelengths due to metal lines \citep{2009bufano}. The effect of line blanketing in the region $<$\,3000~\AA~is much more prominent after +\,13.5~d, and it continues to dominate, with fluxes declining in this region.

The NUV spectrum of SN~2023ixf is compared with a few Type II SNe such as ASASSN-15oz \citep{2019MNRAS.485.5120B}, SN~2017eaw \citep{2019ApJ...876...19S}, and SN~2021yja \citep{2022ApJ...934..134V} at similar epochs in the bottom-right panel of Figure~\ref{fig:UVSpectra}. Two spectra of SN~2021yja (+\,9~d and +\,14~d) are from HST. All other spectra used for comparisons are from Swift/UVOT. Initially, the UV spectra of Type IIP SNe were thought to be homogeneous \citep{2008ApJ...685L.117G}, but as the number grew, the dissimilarities became more evident \citep{2023arXiv230501654B, 2023arXiv230406147V}. 
The absorption feature around 2700~\AA\ arising from \ion{Mg}{2} is observed in all the SNe. The feature around 2900~\AA\ was observed in SN~2023ixf, SN~2017eaw (IIL) \citep{2019ApJ...876...19S}, SN~2022wsp (IIP) \citep{2023arXiv230406147V} and SN~2022acko (IIP) \citep{2023arXiv230501654B}. Detailed modeling for SN~2022acko revealed it to be an absorption window from the close-by \ion{Fe}{2}, \ion{Cr}{2}, and \ion{Ti}{2} absorption complexes \citep{2023arXiv230501654B}. This absorption feature is also observed in SN~2021yja in the spectrum of +\,14~d.

The shape of the continuum is very similar prior to +\,10~d for SN~2021yja and SN~2023ixf. As the spectra evolve, a sharp cutoff in flux $<3000$~\AA\ could be observed beyond +\,10~d in all the SNe compared, indicating a significant line blanketing. Around +\,14~d, the differences in spectra are very apparent, especially in ASASSN-15oz, where in the spectrum below 2700~\AA, we find strong emissions/absorptions, whereas others are devoid of flux comparable to regions beyond 2700~\AA. Slightly higher flux beyond 3000~\AA\ could indicate ongoing interaction \citep{2022ApJ...934..134V}. More SNe need to be observed in the UV, specifically within the first three weeks of the explosion. This will be crucial in understanding the progenitor characteristics, its environment, and its effects on the early evolution and will aid in testing homogeneity in their spectra \citep{2021arXiv211115608K, 2023arXiv230501654B}.

\section{Light Curve Analysis}
\label{sec:lightcurves}

\begin{figure*}[hbt!]
	 \resizebox{0.50\hsize}{!}{\includegraphics{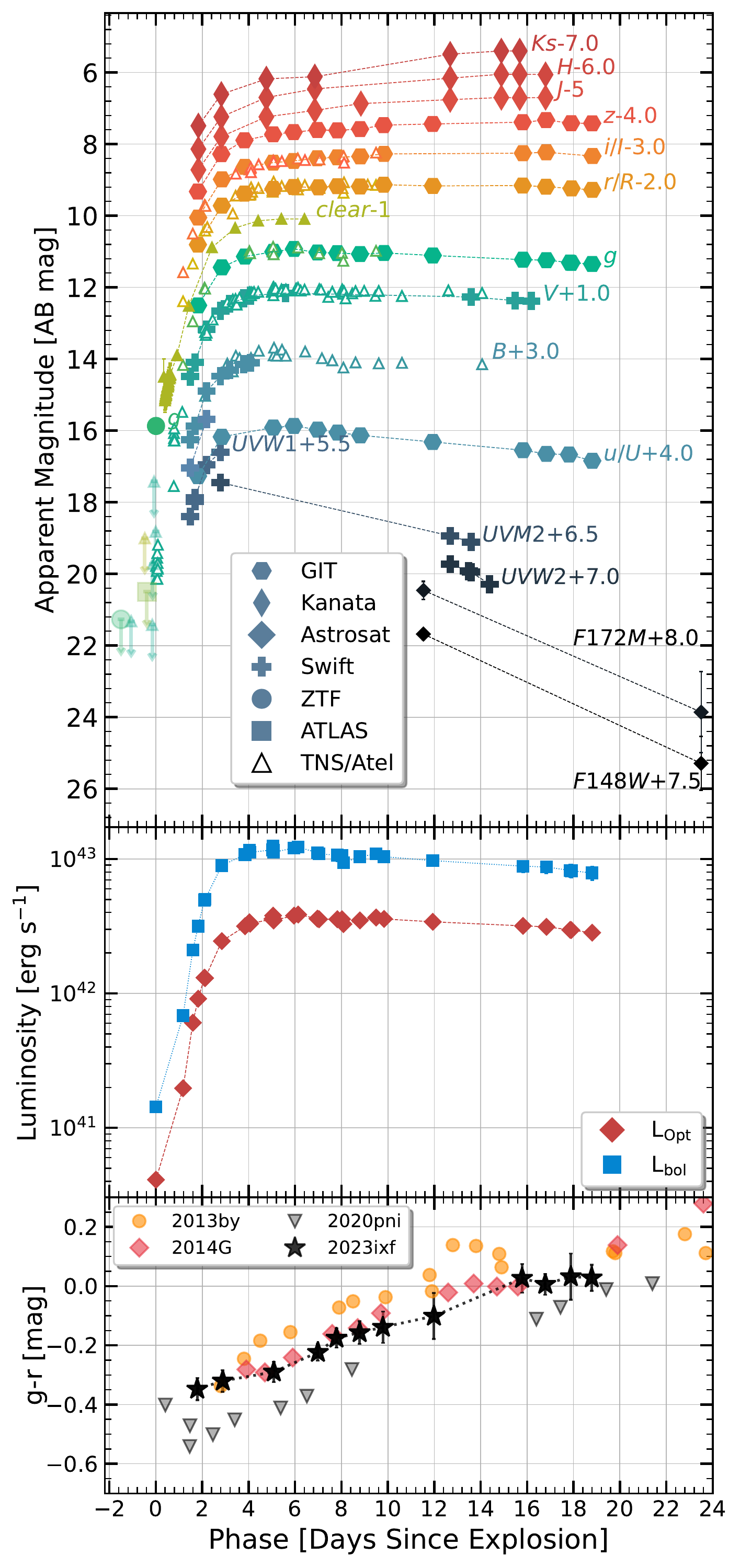}}
   \resizebox{0.50\hsize}{!}{\includegraphics{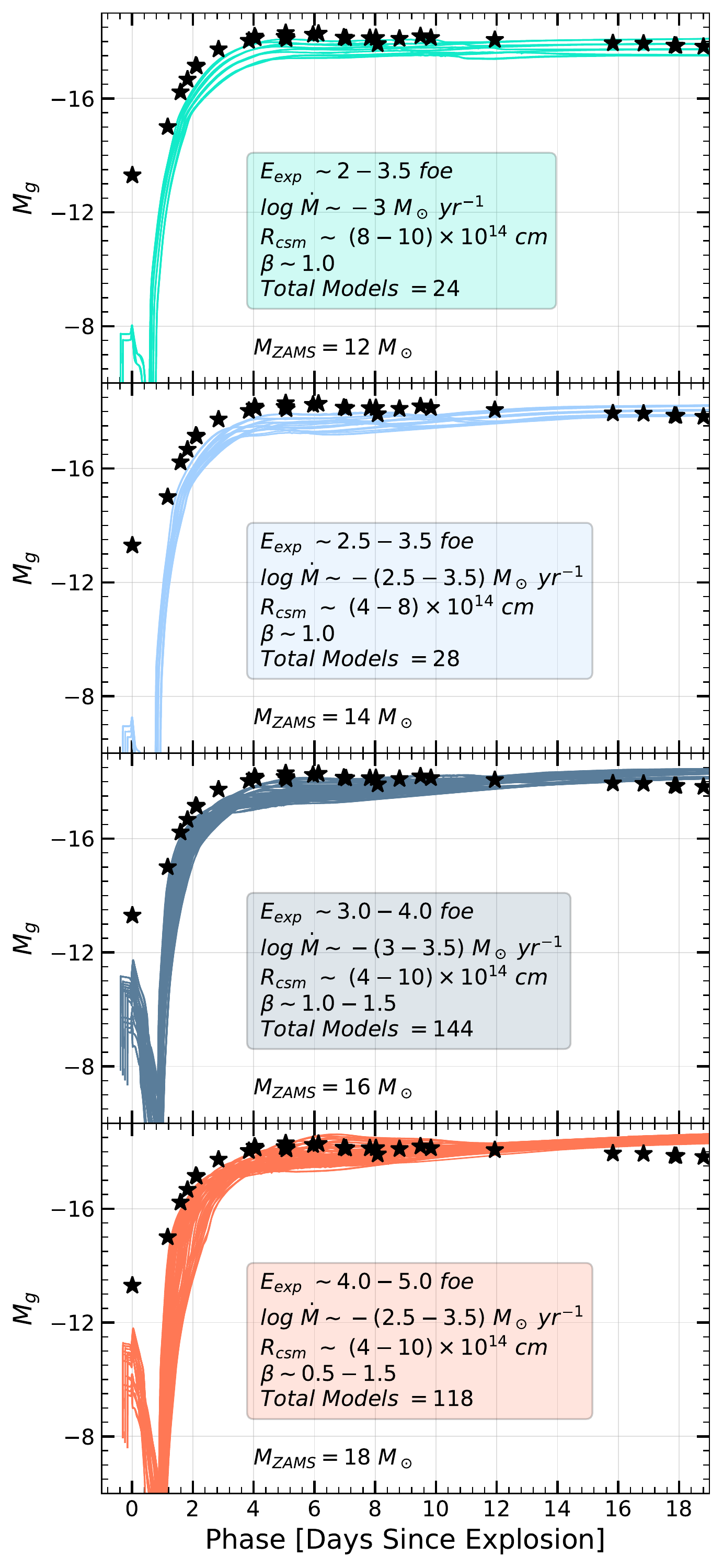}}
  
    \caption{{\it Left}: (Top to Bottom) Multiband photometry is shown along with the data compiled from public sources. The second plot shows the bolometric light curve evolution with the obtained black body parameters. The bottom plot shows the color evolution of SN~2023ixf along with the other SNe with observed flash features. {\it Right:} Best model light curves that could represent the g-band light curve evolution of SN~2023ixf obtained out from a large sample of $>$170,000 models presented in \citet{2023PASJ...75..634M} for different progenitor masses. The photometric data obtained using AstroSat, GIT, Kanata, and Swift are available as data behind the figure.}
    \label{fig:LightCurveAnalysis}
\end{figure*}

The multiband light curves based on observations from the various facilities described in Section~\ref{sec:obsdata} are shown in Figure~\ref{fig:LightCurveAnalysis}. We converted all the pre and post-discovery public data to AB magnitude scale and included it with our dataset using the transformations described in \citet{2007AJ....133..734B}. The public dataset reported is very helpful in putting tight constraints on the explosion epoch \citep{2023arXiv230606097H}.  

We do not see the fast declining phase after maximum, the $s1$ phase, usually attributed to the initial cooling phase post-breakout \citep{2014Anderson}, in the $V$-band light curve of SN~2023ixf. Instead, it declines very slowly for the initial few days, at $\rm 1.18^{+0.49}_{-0.51}~\rm{mag}~100~d^{-1}$, right after it reached a peak $V-$band magnitude of $-18.06\pm0.07$~mag around $\sim 5$~d after explosion. The peak magnitude falls at the brighter end of Type II SNe. The peak V-band brightness is comparable with SN~2013by \citep{Valenti2015_2013by} and SN~2014G \citep{2014GTerraran}, which were classified as Type IIL, although with many similarities to the Type IIP sub-class. SN~2014G also showed flash ionization features in early spectral evolution. While the initial decline of SN~2023ixf is inconsistent with that of Type IIL, its evolution at later phases is yet to be probed. Although the early spectra indicate interaction with a nearby dense CSM, SN~2023ixf is not extremely bright in the UV bands like Type IIn SNe. 

The observed rise time of $\sim 4-5$~d is shorter than other normal Type II SNe, which, on average, take $\sim$ 10 days to reach the peak \citep{2016MNRAS.459.3939Valenti}. We compare the $g-r$ color with similar events that showed flash features such as SN~2013by \citep{Valenti2015_2013by, Black2017_2013by}, SN~2014G \citep{2014GTerraran}, and the bluest Type II SN~2020pni \citep{2022ApJ...926...20T}. The color evolution is similar to these events for the initial $\sim$\,20 d but slightly redder than SN~2020pni. The NIR light curves are also presented in \citet{2023arXiv230600263Y} up to a week post-explosion. We show the evolution beyond that and observe that the flux increases in the NIR, possibly due to pre-existing dust around the ejecta. The presence of pre-SN dust is also described in \citet{2023arXiv230606162N}. 

The early prolonged flash features indicated the presence of a dense CSM around the progenitor. Recently, \citet{2023PASJ...75..634M} provided a comprehensive set of grids for model light curves that could shed light on the structure of CSM and its effects on the early light curve of interacting Type II SNe. In their work, a confined CSM is attached over radius, $R_0$, for five progenitors with mass ranging from 10 to 18 M$_\odot$. The CSM density structure follows from \citet{2018MNRAS.476.2840M}, whereas the wind velocity, $v_{wind}$ at a distance $r$ was taken to be in the form as given below:
\begin{equation}
    v_{wind}(r) = v_0+(v_\infty-v_0)\left (1-\frac{R_0}{r}\right )^\beta,
\end{equation}
where $v_0$ and $v_\infty$ are the initial wind velocity at the surface of the progenitor and terminal velocity, respectively, and $\beta$ is a wind structure parameter that determines the efficiency of wind acceleration. 

These model light curves can be used to constrain the very early light curve behavior of Type II SNe. Our work utilizes the well-sampled $g$-band light curve of SN~2023ixf to compare with the model grid of interacting Type II SNe generated by \cite{2023PASJ...75..634M}. We used the models with $\rm ^{56}Ni$ mass in the typical range of 0.01 to 0.04 M$_\odot$ \citep{2014Anderson}. Furthermore, we found that the initial light curves are insensitive to the $\rm ^{56}Ni$ mass. We iterated over each parameter ($\rm E_{exp}$, $\beta$, $R_{CSM}$, and $\dot{M}$) in succession, keeping others fixed with their full range for a single run. This procedure is repeated for 12, 14, 16, and 18~$\rm M_\odot$ progenitor models. We categorically reject models which show significant deviations from the observed light curves based on their peak luminosities and rise times. Subsequently, we do this for other parameters constraining the values for previous parameters. The best-fitting models for each progenitor are shown in Figure~\ref{fig:LightCurveAnalysis}. We note that the slow early rise till day 2 is not captured by any of the models, and the later evolution is such that either the rise or plateau could be matched but not the entire light curve. Since we are concerned about the initial rise, we do not probe it further; detailed hydrodynamical modeling specific to this particular event will be required to understand the entire light curve evolution. Further, the degeneracy in the progenitor masses could not be lifted by these models, but these models give a very tight constraint on the radius of the outer CSM utilizing the rise times of the model light curves. The dense CSM is confined to $\rm 4.0-10.0\times10^{14}~cm$. Further, $\beta$ varies from 0.5 to 1.5 depending on the progenitor mass, which is close to the typical values for RSGs ($\beta >1$). The $\beta<1$ value obtained for $\rm M_{ZAMS}$ would accelerate winds slightly faster and cause less dense CSM in the vicinity, which is not the case for SN~2023ixf. The mass loss rate is also slightly on the higher end ($\rm 10^{-3.0\pm0.5}~M_\odot~yr^{-1}$). The average density of the CSM comes out to be $\rm \sim10^{-14}~g~cm^{-3}$ which is in line with the values obtained in \citet{2023bostroem} but below the values inferred in \citet{2023arXiv230604721J} obtained from the detailed spectral modeling. The mass-loss rates align with the density limits of CSM derived from the non-detection of radio emission (230 GHz) at early times \citep{berger2023}. For a typical RSG ($\sim 500~M_\odot$), the above would translate to a mass loss $\sim 14-18$ years before the explosion. But as seen in \cite{2023arXiv230607964S}, wind speeds measured using high-resolution early spectra are one order higher than what is assumed in the model parameters, which would give an eruptive mass loss timeline to be around 2 years before the explosion. However, wind acceleration cannot be ruled out. Another parameter that is tightly constrained by the models is the explosion energy. Only the models with explosion energies more than 2.0 foe could match the observed $g$-band flux. The explosion energy increases as the progenitor mass is increased. The explosion energy obtained is higher than for the usual Type II SNe.

In a recent work, \citet{2023arXiv230403360K} presented various light curves of transients arising from interacting SN. These include the interaction of SN ejecta with no CSM to a very heavy CSM. Considering the latent space of luminosity and rise-time presented in that work, we find that the light curve evolution of SN~2023ixf (for the period presented in this work) appears to be similar to the model light curves for shock-breakout in a light-CSM scenario. Comparing the rise-times and peak luminosity of SN~2023ixf with the shock-breakout happening inside the CSM, we find that it falls within 0.01\,$<$\,M$_{CSM}$\,[M$_\odot$]\,$<$\,0.1. Using the parameters obtained from light curve analysis, we get a CSM mass ranging from 0.001-0.03~M$_\odot$ (assuming $v_{\rm{wind}}=10$~km~$s^{-1}$), where the upper limit is well within the range obtained from \citet{2023arXiv230403360K}. It indicates the mass loss rate could have been even higher than $\rm 10^{-2.5}~M_\odot~yr^{-1}$, as also being reported in \citet{2023arXiv230604721J, 2023arXiv230703165H}.

\section{Summary}
\label{sec:discussion}

This work presents an extensive set of early-phase observations for the closest CCSN in the last 25 years, SN~2023ixf, that exploded in M~101. The panchromatic observations covered wavelengths from the FUV to NIR regime using both ground and space-based observatories. The multi-band photometry spans FUV to NIR, spanning up to $\sim$\,23 d since the explosion. Light curves were compared with a large model light curves grid to infer nearby dense CSM properties.

Detailed spectral coverage in FUV, NUV, and optical during the first $\sim 25$~days since the explosion is presented, beginning within 2 days from the explosion. The lines due to \ion{Mg}{2}, \ion{Fe}{2} in the NUV, and \ion{C}{3}, \ion{C}{2}, \ion{Si}{4}, \ion{He}{2} in the FUV were identified. The early ($<$\,7 d) spectral sequence of SN~2023ixf indicates the presence of a dense CSM. There are no significant signatures subsequently, except for an intermediate-width emission feature of H$\rm\,\alpha$ after $+7$\,d. The high-resolution spectra presented by \citet{2023arXiv230607964S} show the presence of an intermediate-width P-Cygni profile during this phase, lasting for about a week, arising in the post-shock, swept-up CSM shell. The line profile during the photospheric phase beginning $\sim$\,16 d shows a multi-peaked/boxy profile of H$\rm\,alpha$, indicating an ongoing CSM interaction with a shell-shaped CSM with an inner radius of $\sim$\,75 AU and an outer radius of $\sim$ 140 AU. Considering a standard RSG wind velocity, the progenitor likely experienced enhanced mass-loss $\sim$\,35 - 65 years before the explosion. All the above inferences from our multi-wavelength observations indicate a multi-faceted circumstellar matter around the progenitor of SN~2023ixf. 

The early phase light curve of SN~2023ixf is influenced by the presence of dense nearby CSM, which was likely accumulated due to enhanced mass loss(es) during the later stages of the progenitor's evolution. SN~2023ixf was found to have a very bright peak luminosity ($M_V\approx -18.1$~mag), much higher than the average luminosity for Type II SNe ($M_V\approx-16.7$~mag). Light curves were compared with a large model grid of interacting SNe with varied progenitor masses and CSM properties to infer the properties of the dense CSM in SN~2023ixf. Based on our comparison with light curve models, the high luminosity is likely a mix of interaction with a confined CSM and an inherently energetic explosion. We cannot conclusively decipher the weightage of the above components to the overall luminosity of SN~2023ixf; hence, further monitoring is required. We will continue to carry out the multi-wavelength follow-up of SN~2023ixf.

\section{Software and third party data repository citations} \label{sec:cite}

\vspace{5mm}
\facilities{HCT: 2-m, GIT: 0.7-m, KT: 1.5-m, Swift (UVOT), and AstroSat (UVIT and SXT)}

\software{astropy \citep{astropy:2013, astropy:2018} , emcee \citep{2013PASP..125..306F}, IRAF \citep{93_tody}, HEASoft  \citep{2014ascl.soft08004N}, matplotlib \citep{Hunter:2007}, pandas \citep{mckinney-proc-scipy-2010}, numpy \citep{harris2020array}, scipy \citep{2020SciPy-NMeth}, Jupyter-notebook \citep{Kluyver2016jupyter}, seaborn \citep{Waskom2021}, and SYNAPPS \citep{2011PASP..123..237T} }. 

\section*{acknowledgments}

We thank the anonymous referee for an in-depth review that helped improve the manuscript. RST thanks Sergiy S. Vasylyev for providing HST spectra of SN~2021yja. The GROWTH India Telescope (GIT) is a 70-cm telescope with a 0.7-degree field of view, set up by the Indian Institute of Astrophysics (IIA) and the Indian Institute of Technology Bombay (IITB) with funding from  Indo-US Science and Technology Forum and the Science and Engineering Research Board, Department of Science and Technology, Government of India. It is located at the Indian Astronomical Observatory (IAO, Hanle). We acknowledge funding by the IITB alumni batch of 1994, which partially supports the operation of the telescope. Telescope technical details are available at https://sites.google.com/view/growthindia/. The HCT observations were made under our accepted ToO proposal HCT-2023-C2-P25. We thank the staff of IAO, Hanle, and CREST, Hosakote, that made these observations possible. The facilities at IAO and CREST are operated by the Indian Institute of Astrophysics, Bangalore. DKS acknowledges the support provided by DST-JSPS under grant number DST/INT/JSPS/P 363/2022. This research has made use of the High-Performance Computing (HPC) resources\footnote{\href{https://www.iiap.res.in/?q=facilities/computing/nova}{https://www.iiap.res.in/?q=facilities/computing/nova}} made available by the Computer Center of the Indian Institute of Astrophysics, Bangalore. This research made use of \textsc{RedPipe}\footnote{\url{https://github.com/sPaMFouR/RedPipe}} \citep{2021redpipe}, an assemblage of data reduction and analysis scripts written by AS. This work uses the SXT and UVIT data from the {\it AstroSat} mission of the Indian Space Research Organisation (ISRO). The SN was observed through multiple ToO proposals, and the data were made public through the ISSDC data archive. We thank the SXT and UVIT payload operation centers for verifying and releasing the data via the ISSDC data archive and providing the necessary software tools. This work has also used software and/or web tools obtained from NASA's High Energy Astrophysics Science Archive Research Center (HEASARC), a service of the Goddard Space Flight Center and the Smithsonian Astrophysical Observatory. This work was also partially supported by a Leverhulme Trust Research Project Grant. This research also made use of the NASA/IPAC Extragalactic Database (NED\footnote{\url{https://ned.ipac.caltech.edu}}), which is funded by the National Aeronautics and Space Administration and operated by the California Institute of Technology.



\bibliography{SN2023ixf}{}
\bibliographystyle{aasjournal}



\end{document}